\documentclass{ws-ijmpd}
\usepackage[super,compress]{cite}
\begin{document}

\markboth{M.\ Spaans}
{Topological Extension of GR: Quantum Space-Time, Dark Energy, Inflation}

%
\catchline{}{}{}{}{}
%

\title{A Topological Extension of General Relativity to Explore the Nature of Quantum Space-Time, Dark Energy and Inflation}

\author{M.\ Spaans}

\address{Kapteyn Astronomical Institute, University of Groningen, P.O.\ Box 800, 9700 AV Groningen, The Netherlands\\
spaans@astro.rug.nl}

\maketitle

\begin{history}
\received{Day Month Year}
\revised{Day Month Year}
\end{history}

\begin{abstract}
General Relativity is extended into the quantum domain. A thought experiment is
explored to derive a specific topological build-up for Planckian space-time.
The presented arguments are inspired by Feynman's path integral
for superposition and Wheeler's quantum foam of Planck mass mini black
holes/wormholes. Paths are fundamental and prime 3-manifolds like $T^3$,
$S^1\times S^2$ and $S^3$ are used to construct quantum space-time.
A physical principle is formulated that causes observed paths to multiply:
It takes one to know one. So topological fluctuations on the Planck scale
take the form of multiple copies of any homeomorphically distinct path
through quantum space-time.

The discrete time equation of motion for this topological quantum
gravity is derived by counting distinct paths globally.
The equation of motion is solved to derive some properties of dark
energy and inflation.
The dark energy density depends linearly on the number
of macroscopic black holes in the universe and is time
dependent in a manner consistent with current astrophysical
observations, having an effective equation of state $w\approx -1.1$ for
redshifts smaller than unity.
Inflation driven by mini black holes proceeds over ${\rm n}\approx 55$
e-foldings, without strong inhomogeneity, a scalar-to-tensor ratio
$r=ln(7)/{\rm n}\approx 0.036$ and a spectral index $n_s=1-r\approx 0.964$.
A discrete time effect visible in the cosmic microwave background is suggested.
\end{abstract}


\ccode{PACS numbers: 04.20.Gz; 98.80.Hw; 04.60.Pp; 95.36.+x; 04.70.Dy}


\section{Introduction}	

General Relativity (GR) is one of the most beautiful and succesful physical
theories.$^{1,2}$ It links the action of gravity to the geometry of
space-time. GR's
extension to the quantum domain has been explored over many years by many
authors, e.g., quantum foam ideas for geometric dynamics, loop quantum
gravity, loop algebras for quantum connectivity in four dimensions, spin foams
that suggest a lower effective dimensionality of space-time and string theory
that requires a much higher dimensionality.$^{3,4,5,6,7,8,9,10,11,12,13}$

Early work in Refs.~3-5 on the Wheeler-deWitt equation already
considered quantum geometric dynamics through a sum over all possible
3-geometries, including topologically non-trivial ones. In this, a proper
measure and factor ordering were identified as problems.
Another line of research that goes beyond pure geometry is loop quantum
gravity. In it, well chosen operators allow one to quantize space-time
geometry and derive important insights into the quantum properties of
BHs.$^{6,7}$
Also, the ideas of Mach, a great motivation for Einstein, are still
powerful today in the form of approaches that highlight the fundamental
importance of background independence.$^{14}$ Indeed, although local
formulations of physical laws have proven to be very succesful, a truly
self-consistent universe appears only to be constructable if one explicitly
incorporates a global notion of space-time using topology.$^{12,13}$

In this context, Wheeler's quantum foam of Planck mass mini black holes (BHs)
or wormholes remains an
intuitively appealing picture of quantum space-time. The lack of conformal
invariance in the Einstein equation leads to the inevitable presence of
large metric fluctuations.
Also, the Schwarschild BH is a perfectly fine solution of the
Einstein equation. This while the evaporation of BHs, as well as the
thermodynamic laws that their horizons obey, makes them the perfect
entities to bridge classical and quantum space-time.$^{15,16}$
Of course, quantum space-time may host more than just mini BHs and likely calls
for a completely new structure altogether.

Nevertheless, the absence of
topology in GR, while it predicts BHs, suggests that the mere continuity
of space-time may hold important clues to quantizing it. Also, if the
unstable nature of the quantum foam could be remedied, i.e., if its
(effective) energy density is lowered in a controlled manner, then this may
help to resolve the discrepancy between the large expected vacuum energy and
the perceived low value of present-day dark energy.$^{17,18,19,20,21}$

Furthermore, topology deals with connectivity while geometry pertains to
shape. The connectivity of space-time, i.e., the existence of
homeomorphically\footnote{A homeomorphism between two topological spaces is
a bijective continuous map with a continuous inverse.}
distinct paths between arbitrary
points $A$ and $B$, lies at the heart of quantum physics. It is the path
integral formulation of Feynman that shows how the superposition principle
is an expression of path multiplicity.$^{22}$ The fact that many of the paths
in Feynman's path integral are continuous but not differentiable favors an
a priori topological approach. Finally, if one views paths as fundamental,
then Nature must have some way to construct them and assert that they are
truly distinct. This is a highly non-trivial task in the presence of the
quantum uncertainty in Planck scale geometry discussed above.$^{3,4}$

This work therefore finds its inspiration in Wheeler's quantum foam and
Feynman's path integral formulation for quantum superposition, and uses
topology as the natural language of quantum space-time. The efforts in
Refs.~12 and 13, which use algebraic topology to define quantum geometry, are
taken as a basis. These are elaborated upon to highlight
the crucial role of a multiply (so non-simply) connected space-time for
quantum gravity, dark energy and inflation.

The section below
presents a thought experiment that provides a physical basis for the
mostly mathematical considerations of Ref.~12. Subsequently, the equation of
motion for the ``topological dynamics'' of quantum gravity is derived, and
solved under certain approximations to explore the nature of dark energy and
its relation to inflation. In all, this paper is self-contained.

\section{Topological Dynamics}

\subsection{Thought Experiment: The Multiplication of Paths}

Consider the following thought experiment for an observer who wishes to
assess the structure of his space-time through some measurement process.
Excluding the philosophically interesting case of a completely isolated
observer, it seems reasonable to assume that the observer can detect
phenomena and come to some quantitative view of his universe through a
measuring apparatus. What properties of space-time should the observer
measure and how can he quantify these?

Going back to the introduction of this paper, it is good to recall the
central tenet of quantum physics: the superposition principle. At the root
of all quantum phenomena lies the notion of space-time paths. It is with
paths that the fabric of space-time can be woven and the superposition
principle of quantum physics be expressed, at the same time.

Most straightforwardly the observer can start by choosing a few (material)
objects and measuring the paths these objects take to map out properties of
space through time. While doing so for smaller and smaller objects and
decreasing scales, he will encounter the distinct paths that quantum particles
travel along. For example, when he performs the well known double slit
experiment and
identifies interference effects because electrons follow a multitude of routes.

At this stage, he can start to quantify the properties of his space-time by
identifying different classes of paths. E.g., one class of paths contains all
those paths that go through slit 1 and another class those that go through
slit 2.
Of course, the two classes of paths are a priori a consequence of matter (the
double slit), but the observer decides to extend his experiments all the way
to the Planck scale to explore the spirit of superposition.

On the Planck scale, geometric fluctuations in space-time are large and
inhibit him to perform accurate measurements.
This does not surprise him because the lack of conformal invariance of the
Einstein equation already implies this quantum behavior.
However, these fluctuations do not
preserve the shape of space-time, while the observer wants to measure what
paths comprise space-time despite any such quantum uncertainty in geometry.

He then recalls the double slit experiment, with the two distinct classes
of paths, and realizes he can still measure paths that cannot be obtained
from each other by simply changing the shape of space. I.e., paths are
detectable if they cannot be transformed into each other through a
homeomorphism of space-time. Consequently, the properties of such paths are
assessable by the observer.

But, when taking a good look at his, by now formidable, measuring apparatus,
it is also obvious to him that he is as much observer as actor in these
experiments. The quantum uncertainty in geometry that plagues him, is as much
caused by his apparatus as it is a consequence of the lack of conformal
invariance in GR.
While performing experiments in this manner, he develops the following insight.
It is through the observational comparison of paths that one comes to an
assessment of their identity, and thus the structure of quantum space-time.
As such, no path can exist, or remain to exits, on its own when scrutinized
and the observer thinks:
``It takes one to know one. So topological fluctuations on the Planck scale
take the form of multiple copies of any homeomorphically distinct path
through quantum space-time.''

So due to Planckian quantum effects, the very act of identification requires
paths to multiply, in 4-space.
Although puzzled, the observer also realizes that the
minimal form that any identification of an entity takes, is to at least count
it. Apparently, his counting of paths is prone to quantum fluctuations with
pertinent consequences for Planckian space-time.

The observer then summarizes his observational acts as follows.
The minimal form of space-time path identification is to count them.
In order to count paths, he must confirm their existence by recognizing their
intrinsic properties. This requires at least a proper example to
observationally compare to
and implies a sense of multiplicity, if he is to make any quantitative sense
of space-time and its quantum geometric uncertainty on the Planck scale.

Being somewhat mathematically inclined, the observer recognizes the importance
of distinction when speaking of paths and their multiplicity. Path distinction
is the expression of homeomorphic inequivalence. I.e., using the symbol $\sim$
for equivalence, paths $X$ and $Y$ that obey $X\sim Y$ belong to the same
class $[X]$ because a continuous map transforms one into the other.
Hence, to assert that there are paths $[X]$ the observer
concludes that he must have at least one other class $[X']$ to compare to.
For example, consider a landscape with a bridge. One can walk along many paths
$[X]$ across the bridge and these paths are all topologically distinct from
those paths $[Z]$ that choose to cross the water. The comparison class $[X']$
then simply represents all those paths crossing another bridge.

Pondering this some further, the observer realizes that he is only one and
that Nature herself needs to take on the task of counting distinct space-time
paths. In doing so unremittingly, the very act of counting distinct paths must
result in pertinent and persistent changes to the structure of scrutinized
quantum space-time.

The observer concludes that the nature of these multiplicative changes
preserves the intrinsic topological properties of the original paths.
That is, quantum
fluctuations in geometry are merely incidental on the Planck scale. It is the
connectivity of the bridge mentioned above that matters, not its shape.
Furthermore, when counting that bridge, the interaction between observer and
observee requires Nature to build at least one other bridge.
If Nature would not, then no proper confirmation of a path's identity is
possible at all. Because first cause lies in the act of identification
(counting), path multiplication is an expression of observation.

In everyday life such a notion appears untenable, but on the Planck scale it
has long been argued$^{3,4}$ that any formal measurement on the
order of a Planck time, using a Planck mass worth of energy, must lead to a
space-time with topologically exotic objects like wormholes, loops, etc.
I.e., Planckian space-time is multiply connected, supporting a plethora of
homeomorphically distinct paths that are caused by the act of observation.

This space-time path multiplication phenomenon, first discussed for local
paths in Ref.~12 and referred to as induction for BHs in Ref.~13, is elevated
here to a fundamental physical property of quantum space-time:
the multiplicity principle (MP). It is formulated as follows.
\medskip

{\it Paths that are distinct under the continuous deformation of
space-time constitute a fundamental observable property of space-time.
To confirm the identity of such paths on the Planck scale, they must at
least be countable.
Under the observational act of counting, no distinct path can (remain to) exist
individually. If it would, then its intrinsic properties would be unknown
and its fundamental nature undefined. Instead, it takes one to know one.
So, when scrutinized by any observer,
these distinct paths must multiply and thus build quantum space-time.
Therefore, topological fluctuations on the Planck scale take the form of
multiple copies of any homeomorphically distinct path through quantum
space-time.}

\medskip
The above considerations are in line with usual quantum phenomenology, in
which no formal distinction can be made between observer and observee, and
interactions between them affect both in a pertinent manner.
The MP allows Nature to distinguish paths, despite any quantum uncertainty
in geometry, using topology. The MP also forces Nature to treat paths as
{\it dynamical} entities on the Planck scale, requiring multiplication.
The information on the (number of) different equivalence classes of paths
is distributed
through homeomorphisms. This information therefore constitutes a global
quantity. I.e., topology cares only for integer numbers, and a Machian
perspective on the universe results.

\subsection{Quantum Space-Time Structure}

The MP can be reworded as that one single path modulo homeomorphisms has the
same physical significance as no path. Or equivalently, that the
minimal$^{12}$ building block of space-time is the loop $S^1$.
That is, a loop is the combination of two
paths from A to B such that one cannot be deformed into the other by a
homeomorphism.

\subsubsection{The Prime Manifolds $T^3$, $S^1\times S^2$ and $S^3$}

When dealing with space-time, so with 4 dimensions, the natural building
blocks are so-called prime 3-manifolds and their number evolution in time.
Any topologically non-trivial space-time manifold$^{23,24,25}$ can be
constructed by adding together 3-primes, embedded
in 4 dimensions, through a connected sum. These
3-primes (e.g., handles, three-tori, three-spheres) lend their name from the
property that they themselves cannot be written as the connected sum of
smaller 3-dimensional units. Furthermore, interactions among
prime manifolds can never change their intrinsic properties, like the
number of closed loops they contain.

The MP then dictates that the path dynamics of any 4-dimensional quantum
space-time must be carried by 3-primes that contain {\it pairs} of paths
through their $S^1$ loop make-up. Once a basis set of 3-primes is chosen,
the MP also imposes that quantum space-time is constructed by them.
In Refs.~12 and 13 a 3-prime basis set has been proposed that consists of the
three-torus $T^3$, the handle $S^1\times S^2$ and the three-sphere $S^3$.
In this work, somewhat different arguments are presented but the same basis
set of 3-primes is found.

\medskip
1) A multiply connected space-time should be locally flat and locally
isotropic in
the absence of gravity. Using the $S^1$ loop, there is only one 3-prime
that is locally isotropic and non-simply connected. This is the
three-torus$^{23}$,
$T^3=S^1\times S^1\times S^1$. The three loops that $T^3$ contains require
it to be embedded in four dimensions. Indeed, the three loops are topological
identifications, through the fourth dimension, of the front-back, top-bottom
and left-right surfaces that form the boundaries of the solid cube $I^3$.

When confined to the Planck scale, the three-torus also introduces, in a local
sense, the homeomorphically distinct paths that are needed for the
superposition principle as expressed by Feynman's path integral.$^{22}$
Note that the Euler characteristic of $T^3$ is zero, so it is the boundary
of a Lorentz 4-manifold.

\medskip
2) The Einstein equation enjoys the Schwarzschild solution for a macroscopic
BH. The latter constitutes the low energy limit of Wheeler's quantum foam.
I.e., Planckian BHs are transient under Hawking evaporation while macroscopic
BHs are among the most stable entities in the universe on entropic grounds.
One should note then that the loop containing handle $S^1\times S^2$ is just
like the Schwarzschild BH solution when one considers Hawking evaporation in
4-space, as follows. 

Even a macroscopic BH evaporates very slowly. So, in a topological sense,
the Schwarzschild BH\footnote{The same holds for more general BH solutions
like the Kerr metric.} allows a connection between different times for
accreted versus radiated mass-energy. It is therefore topologically
equivalent to a handle $S^1\times S^2$, which connects different spatial
regions, when one treats a macroscopic BH as a time-dependent embedding in
4-space. One could consider the handle or BH as a bridge between entirely
different universes, with orthogonal sets of dimensions. However, a single
multiply connected 4-space is considered to be the universe here, albeit with
many superposed quantum histories (see further in section 2.2.3).

In this work, there is no limitation imposed by the null energy
condition. Planckian wormholes evaporate on the order of a Planck time.
Macroscopic BHs only possess a wormhole topology across cosmically separated
time slices (probably in a completely incoherent fashion).
Obviously, a countable quantity like the number of BH event horizons is
invariant under homeomorphisms of four-dimensional space-time.
Note that the Euler characteristic of $S^1\times S^2$ is zero, so that it
is also the boundary of a Lorentz 4-manifold.

\medskip
3) When one counts pairs of distinct paths, there should be the possibility of
finding zero. With the loop being the central object of a multiply connected
space-time, the zero represents the absence of loops and thus a simply connected
space-time. The latter is equivalent to the three-sphere $S^3$ and
homeomorpisms thereof. Diffeomorphisms are a subset of all homeomorphisms.
The coordinate invariance of shape should therefore appear as one goes to
large $S^3$ scales (see further in section 4.2).

\medskip
One can then conclude that a plausible basis set consists of the
non-chiral prime 3-manifolds $T^3$, $S^1\times S^2$ and $S^3$, where 1) the
three-torus is confined to the Planck scale and carries local isotropy and
superposition for a multiply connected space-time; 2) the non-irreducible
handle appears as the large scale solution of the Einstein equation, i.e.,
the Schwarzschild BH;
3) the topologically trivial three-sphere represents the zero loop space-time,
i.e., the unit element for path multiplication.

\subsubsection{Construction through the Connected Sum}

With the basis set in place, it is necessary to apply the MP.
Starting with $T^3$, a space-time with a single three-torus
cannot exist by itself. Rather, the act of counting multiplies it. Even though
embedded in 4-space, $T^3$ is a 3-prime and its paths in 3-space
proceed through the six faces that it possesses. Hence, the multiplication
of three-tori that the MP requires can only connect the additional three-tori
through 3-dimensional junctions (homeomorphic to the solid cube).
In other words, one has a lattice.

Such a lattice of three-tori can be
constructed by performing a connected sum. I.e., by using three-ball
surgery to connect each face of a three-torus to a face of another
three-torus through a junction. So a lattice $L$ of three-tori can be
written as

\begin{equation}
L(T^3) = \oplus^i T^3_i,
\end{equation}
where $\oplus$ denotes a connected sum. The Einstein summation convention
is adopted and $i$ can be arbitrarily large unless noted otherwise.
It is important to realize that the origin of $\oplus^i T^3_i$ in Eq.~(1)
lies purely in the topological dynamical consequences that application of the
MP has. That is, only connected pairs of three-tori have comparative meaning.

The construction of the lattice can be performed along each of the 6
directions defined by the 6 faces. Subsequently, one can always apply
a homeomorphism $H$ that relocates the junctions between three-tori.
Hence, for $i>>1$ and all homeomorphisms $H_k\{L(T^3)\}$, the equivalence
classes $[H_k\{L(T^3)\}]_p$ contain all distinct paths $p$ on $L(T^3)$.
Einstein gravity that bends $L(T^3)$ on large scales is then naturally
allowed by a subset of smooth homeomorphisms.

Through $L(T^3)$, the motion of matter, like a particle world line, derives
from a superposition of wave amplitudes along many different paths.
One thus obtains, from the MP and expressed by the topology of space-time,
the superposition principle as formulated by the Feynman path integral.
Many of the paths that enter the Feynman path integral are continuous,
but not necessarily differentiable. This is in the spirit of the homeomorphic
invariance that underlies the approach in this work.
The constructed distinct paths are now expressed, independent of quantum
uncertainty in geometry, by the {\it global} multiply
connected character of quantum space-time.

Furthermore, handles can be attached to the lattice of three-tori through
the connected sum as well. This structure is denoted by $L(T^3)^+$, i.e.,

\begin{equation}
L(T^3)^+ = L(T^3) \oplus\ [\oplus^j (S^1\times S^2)_j],
\end{equation}
for arbitrary $j$. Because of the homeomorphic freedom of the connected
sum, handles can connect to three-tori as well as other handles.
The simply connected three-sphere $S^3$ embodies the freedom of
homeomorphisms and one has that

\begin{equation}
T^3 \oplus S^3 \sim T^3,
\end{equation}
and
\begin{equation}
S^1\times S^2 \oplus S^3 \sim S^1\times S^2.
\end{equation}

Even with the heuristic application of the MP, it is therefore possible to
construct a
continuous quantum space-time from 3-primes that possesses, dynamically active,
global paths based on connected $S^1$ loops.
Information on the number of connected three-tori and handles is distributed
through homeomorphisms, as already mentioned above.
That is, the number of three-tori and the number of BHs are global, Machian,
quantities. Knowledge of these dimensionless integer numbers is accessible to
all Planckian observers. This while acquisition of such information proceeds
without violation of causality because homeomorphisms allow any two quantum
space-time points to be infinitesimally close. The latter point deserves
further elaboration.

\subsubsection{The Wave Function of the Universe}

One may wonder how an observer can physically know about the
formation of a BH far-away from him. The essence lies in being an observer
on the Planck scale, so that the equivalence classes $[H_k\{L(T^3)^+\}]_p$
are fully available. Then note the following.

In classical 4-space, one can see past events and their future consequences
combined into one. In quantum space-time this is augmented with the notion of
superposition. Taking a global perspective on quantum events then leads to
a wave function of the universe $\psi$, and one that enjoys correlations
spanning across space and time.
On macroscopic scales, every day experience
tells us that classical processes tend to dominate over quantum correlated
ones. However, on the Planck scale the converse is true.

Hence, the formation of a BH involves the irreducible collapse of $\psi$ to
a state with one extra BH. The universe as a whole, and all Planckian
observers in it, then occupies a new state in the
present. This while before, the universe was in a superposition of states
that involved both the succesful and unsuccesful creation of the BH.

As long as all the chains of events, all the space-time paths $p$, that lead
to a
particular present can be distinguished homeomorphically on $L(T^3)^+$, then
the collapse of $\psi$ is a {\it global} event. Specifically, it is one that
is global for all Planckian observers in terms of countable topological
quantities like BHs, three-tori and the paths through them. The nature of
the information involved in the formation of a BH (just a single integer
number) is thus crucial here. This while the distribution of this information
is an integer change within $\psi$.

There is a clear analogy here with the superposition of two electrons, in
a joint up/down spin state, that are moved macroscopically far apart.
Observation of one electron in an up state leads to wave function collapse
and the instantaneous implementation of the down state for the other electron,
and vice versa. The reason is that no information needs to be exchanged
causally upon wave function collapse. This has already been taken care of
while creating the macroscopic superposition in the first place.

Similarly, the formation of a BH is a history, one of many, that ends with
a quantum mechanically leaky event horizon. In 3-space, at some moment in
time, such an event causes local geometric disturbances that travel causally
from its source. In
4-space, the formation of a handle constitutes a topological change
that globally distinguishes one time-like slice through 4-space from another,
by a {\it single} bit of information.

Indeed, experiments similar to the double slit experiment mentioned above
appear
not to produce an interference pattern if a single bit of information labels
a path$^{26}$, and Schr\"odinger's entanglement is destroyed.
Analogously, BH formation leads to the selection and distinction of (and thus a
choice for) a particular path through 4-space.
These points will be returned to in the following sections, but the
qualitative argument is as stated here.

\subsection{Intermezzo: The $S^1$ Loop in Explorations of Quantum Gravity}

In a phenomenological sense, the importance of loops is of course well
recognized in other fields as well, e.g., loop quantum gravity
and the theory of closed strings.$^{6,7,8,9,10}$ To place the explorations
of this work in context, one can note the following.

\subsubsection{Loop Quantum Gravity}

The main motivation for loop quantum gravity, in terms of physical
perspective, lies in the determination of space-time curvature.
The latter is most elegantly defined through a closed path integration,
which brings out curvature as the differences between parallel transported
vectors.$^{25}$
When the loop is elevated to a fundamental geometric object, one can
succesfully quantize geometry and derive important properties on the
spectrum of the BH event horizon. Also, loop quantum cosmology appears
to be free of a big bang singularity.
In this, invariance under diffeomorphsms is adhered to as motivated by GR.

Loop quantum gravity and topological dynamics share their appreciation for,
and implementation of, background independence.
The main difference between them lies in the topological nature of the MP.
I.e., in the global characterization of quantum space-time using
homeomorphisms and the multiplication of (pairs of) paths as the operative
aspect of quantization.

\subsubsection{String Theory}

The mathematics underlying string theory are beautiful. The 2-spin
excitation that the string possesses as well as the AdS correspondence
are very promising.
Nevertheless, any string (closed or open) inhabits a higher-dimensional
background space-time that needs to be provided for a priori. Even though the
mathematical necessity for such an embedding space is clear, and driven by
anomaly cancellation, the origin and evolution of these extra
dimensions lies beyond the physical principles of string theory.

In string theory, the $n$-fold torus $T_n$ is the connected sum of two-tori
$T^2=S^1\times S^1$. So, for $i=1$ to $i=n$,

\begin{equation}
T_n = \oplus^i T^2_i.
\end{equation}

The n-fold torus is the result of closed string interactions because
individual $S^1$ loops
map out cylinders that build up $T_n$ when glued together. Such a structure
nicely removes the single point interactions that plague particle field
theory. Like $L(T^3)$, the manifold $T_n$ is multiply connected.
That being said, the difference in dimensionality between $T_n$ and $L(T^3)$
is fundamental. While the MP helps one to build a quantum space-time, $T_n$
describes string interactions and not space-time itself.

However, one can identify an interesting connection between string theory and
topological dynamics. If $\partial$ is the boundary operator$^{25}$ with the
usual property
\begin{equation}
\partial^2 = 0,
\end{equation}
then
\begin{equation}
\partial T^3 = 0\ \ \& \ \ \partial T^2 = 0.
\end{equation}
Hence, ignoring boundary terms and noting that $\partial D^2 = S^1$ for the
solid disk $D^2$, one has that

\begin{equation}
L(T^3) = \oplus^i [T^2\times\partial D^2]_i.
\end{equation}

Now imagine that the two topologically identified spatial dimensions of $T^2$
are much larger, by a factor of $n$ compared to the Planck length, than the
third $S^1$ boundary of $D^2$. The latter then
represents much higher energies and constitutes a rapidly varying phase that
stores its effects in the membrane $D^2$. Because of the large metric
fluctuations that occur on the Planck scale, it is natural to identify this
$\partial D^2$ with the metric component of Wheeler's quantum foam.

That is, for the analysis in this section the Planck size surface $D^2$ is
assumed not to be transformed into an event horizon $S^2$, which would require
handles to be attached to $L(T^3)$. Under these conditions,
for $n^2>>1$ but small compared to the total number of three-tori in
$L(T^3)$, one finds locally for $i=1$ to $i=n^2$ that

\begin{equation}
L(T^3) \approx [\oplus^i T^2_i]/n^2 \times \partial D^2 = \frac{T_n}{n^2} \times \partial D^2.
\end{equation}

The sub-lattice $\oplus^i T^2_i$ has a weight $1/n^2$ because there are
two independent directions in $T_n$ (it is a surface) that are each sampled
with a duty cycle of $1/n$ relative to $\partial D^2$. The occurrence of a
weight as part of the connected sum in Eq.~9 may seem odd. However, this
weight simply counts the relative multiplicity with which components are
connected. In Eq.~(9), one has a ratio of $1/n^2$ $T_n$ to 1 $\partial D^2$.
So only 1 of $n^2$ loops in $T_n$ is connected to $\partial D^2$.
The above allows one to properly work, despite homeomorphic invariance, with
fractions of otherwise integer topological quantities.

The above brings out $T_n$ as a lower-dimensional large scale topological
perturbation for $n^2>>1$. I.e., $T_n$ is a slowly varying amplitude of
$L(T^3)$ and one that only weakly affects the rapid phase changes carried
by the surface $D^2$. This because a Planckian observer on $\partial D^2$ can
only count a (small) $1/n^2$ fraction of the total number of closed string
interactions on $T_n$ during a Planck time.

In all, the heuristic value of the $S^1$ loop is beyond doubt. It is the
physical principle guiding its use as a mathematical building block that
sets at least a few of the various approaches to quantum gravity apart.

\section{An Equation of Motion for the Evolution of Quantum Space-time}

So far, the discussion has been mostly heuristic.
To make quantitative progress, an equation of motion should be derived for
the evolution of quantum space-time and it must be solved to make falsifiable
predictions.
In Ref.~12 a mathematical derivation is presented of a quantum space-time
equation of motion, employing loop algebraic topology. A more physical
derivation is presented here.

\subsection{Physical Derivation: Counting Globally Distinct Paths}

The key in the derivation of the equation of motion is 1) to incorporate the
MP through proper counting of paths that are embodied by the basis set of
3-primes; and 2) to embrace the discrete nature of a topologically based
quantum space-time all the way to time itself.

\subsubsection{Right Hand Side}

First, one recalls that basic connectivity, i.e., the need to travel from A
to B, requires homeomorphically distinct paths. The minimal
duo being the loop $S^1$. Second, one reasons that the existence
of the loop-containing prime 3-manifolds $T^3$ and $S^1\times S^2$ follows
directly from Feynman (superposition) and Wheeler (quantum foam), respectively.
Third, one counts all loops in 3-primes self-consistently using the MP,
every Planck time.

An observer, or rather Nature, who wishes to identify a three-torus can do
so while approaching a $T^3$ along any of 6 spatial directions (top,
bottom; front, back; left, right). In order to assess the properties of the
three-torus from some side, within a Planck time, the MP requires an
additional three-torus. The same
holds for all neighboring directions, also within a Planck time.
Combined, this implies that the complete
identification of a three-torus under the MP requires 6 additional copies, and
thus a multiplication factor of 7.

For the handle, the same logic applies. Albeit that a handle constitutes
a bridge and counting can proceed only on either side of the bridge. This
yields 2 sides, spatially separated for a handle and temporally separated
for a slowly evaporating BH, from which to assess the nature of the handle/BH.
The MP then imposes two additional copies, every Planck time,
and thus a multiplication by a factor of 3.\footnote{
For a BH picture, as described earlier, one also finds 2
directions because there is a mass-energy flow going in due to
accretion and one going out due to Hawking evaporation. Here the 2
directions are separated by time rather than space, but in a 4-dimensional
topological manifold that distinction is irrelevant.}

One can formulate the numerical considerations above in terms of the action
of the differential counting operators $\delta_i$, with $i=0$ for the
three-torus, $i=1$ for the handle and $i=2$ for the three-sphere, as

\begin{equation}
\delta_i P_i = \mu_i P_i,
\end{equation}
for the 3-prime manifolds $P_i$, with
\begin{equation}
\mu_0=6\ \ \& \ \ \mu_1=2\ \ \&\ \ \mu_2=0.
\end{equation}

Therefore, a single BH/three-torus becomes a triplet/heptaplet of
BHs/three-tori under the MP. This confirms the need for a lattice of
three-tori under the MP. However, BHs carry mass and are subject to Hawking
evaporation. So, the MP imposes additional pairs of BHs with a success
that depends on the longevity of existing BHs.

\subsubsection{Left Hand Side}

For the left hand side of the equation of motion one first realizes that the MP
deals with global properties, basically just countable numbers. Furthermore,
as mentioned before, homeomorphisms acting on $L(T^3)^+$ allow for causally
close proximity between any two space-time points.
This is not to say that space-time is static, but rather that only a 1st
order in time differential is needed for the left hand side, as with the
Schr\"odinger equation. Furthermore, the nature of the time variable (which
will be addressed in more detail below) has to be taken as discrete.

Afterall, even though the mathematical basis of quantum space-time derives
from continuity, physically measurable quantities that are globally accessible
to a Planckian observer are integer numbers only.
Hence, when speaking of space,
one does so in units of the Planck volume $l_P^3$ for the Planck length
$l_P\approx 1.6\times 10^{-33}$ cm. While for time the proper unit is the
Planck time $t_P\approx 5.4\times 10^{-44}$ s.

When counting as the MP
requires, the massless three-torus has its 4-space dimensions specified by
$l_P$ and $t_P$. However, the handle carries mass. Therefore, the natural unit
of mass for a handle/BH that results from multiplication in the equation of
motion is the Planck mass $m_P\approx 2.2 \times 10^{-5}$ g.
The counted numbers of three-tori and handles are denoted by the
multiplicities $n_i$ and one has that

\begin{equation}
\Delta n_i P_i = n_i(m+1) P_i - n_i(m) P_i,
\end{equation}
for discrete times $(m+1)t_P$ and $mt_P$, in physical units, and integer steps
$m=0,1,2,...\ $.

\subsubsection{Final Form: Discrete Time Evolution of 3-Prime Multiplicities}

The correct connection between the left hand side and right hand side again
involves the MP. Evolution is simply $\Delta n_i$. However for the counting
of 3-primes on the right hand side only a single three-torus and handle have
been considered, with $\mu_0=6$ and $\mu_1=2$. Of course, the act of counting
is performed unremittingly. Therefore, the total multiplication that one
incurs under the MP is $\delta_i n_i P_i$. This yields

\begin{equation}
\Delta n_i P_i = \delta_i n_i P_i = \mu_i n_i P_i.
\end{equation}
Because the index $i$ already labels the 3-primes $P_i$ on which the operators
act, one can also suppress the appearance of $P_i$ and write

\begin{equation}
\Delta n_i = \delta_i n_i = \mu_i n_i.
\end{equation}
Note that there is no topologically dynamical role for the basis 3-prime
$S^3$, as would be expected since

\begin{equation}
\Delta n_2 S^3 = \delta_2 n_2 S^3 = 0.
\end{equation}

However, for the handles/BHs there
are matter degrees of freedom that play a role as well. Afterall, the number
of BHs changes as BHs form by gravitational collapse, evaporate by quantum
processes or merge when brought into close proximity.

These processes are
denoted by $F$ (formation), $E$ (evaporation) and $M$ (merging), respectively.
Because such effects change the multiplicity $n_1$, they are also subject to
the discrete time variable $m$
and written as $\Delta F$, $\Delta E$ and $\Delta M$.
Going from one Planck time to the next, and adding the matter degrees of
freedom terms to the right hand side, one thus finds the quantum space-time
equation of motion as

\begin{equation}
\Delta n_i=\delta_i n_i + \delta_{1i}(\Delta F - \Delta E - \Delta M),
\end{equation}
with $\delta_{1i}$ the usual Kronecker delta. Equivalently, explicitly
including discrete time dependence as well as the masses that BHs have,
one has

\begin{equation}
n_i(m+1) = (\mu_i + 1) n_i(m) + \delta_{1i}(\Delta F(m+)-\Delta E(m+)-\Delta M(m+)),
\end{equation}
where $(m+)$ denotes the change $\Delta$ during the continuous time interval
$(m,m+1)$. Proper accounting of $n_1(m)$ under formation, evaporation and
merging is assumed.
I.e., $\Delta F=1$ means the formation of one BH (hence the plus sign).
Also, $\Delta E=1$ means that one BH has evaporated during one time step
(hence the minus sign), while $\Delta M=1$ means that two BHs have merged
into one (minus sign). Naked singularities may in principle form and
these could yield $\Delta E=0$ after an evaporation event. Gravitational
lensing observations may be used to confirm the possible existence of naked
singularities.$^{16}$

It should be noted, and this is discussed further below, that $n_1(m)$
pertains to BHs of all masses, from the Planck mass to Solar masses and more.
This is particularly relevant for the $\Delta E(m+)$ term.
So one can best write $n_1(m)[\{M_k\}]$ (and also include charge and angular
momentum if they play a role) for the masses $M_k$ that BHs $k=1...n_1(m)$
have. For ease of writing, this will typically be suppressed and $n_1$ is used.

\subsubsection{GR and the Equation of Motion}

Eq.~(17) appears quite simple, but this is slightly
misleading because the matter degrees of freedom term after the Kronecker
delta affects the number of handles $n_1$ in a manner that potentially
involves the full impact of GR.

Indeed, the evolution of the handle multiplicity $n_1$ is in part determined
by the formation, evaporation and merging of black holes. These processes
depend on the cosmic evolution of the energy-momentum tensor and space-time
curvature tensor in the Einstein equation. Afterall, the occurrence of
structure in the universe is intimately tied to the formation of massive
stars, their BH stellar remnants, the subsequent dynamical interactions
between those BHs, and their accretion histories.

Therefore, the longevity and transience of all BHs comes
into play through the right hand side of Eq.~(17). As such, ``extension''
appears to be the correct terminology for the topological approach pursued
in this work. Eq.~(17) serves GR, as follows.

GR remains completely valid down to the Planck scale as long as care is
taken to include the effects of induced BHs and BH evaporation into the
Einstein equation. These effects are determined in sections
5.1 (dark energy and accelaration today),
6.2 (dark energy and inflation in the early universe)
6.3 (amplitude of mass-energy fluctuations) and
6.5 (scaling properties of mass-energy fluctuations).
When included, then the terms $\Delta F$, $\Delta E$ and $\Delta M$ can
be computed as one would otherwise do\footnote{
The computation of the resulting spectrum for BH evaporation knows some
subtleties, but for Eq.~(17) only the overall decrease in mass is relevant.}
when following the evolution of energy-momentum and curvature.

It will turn out that these corrections leave the form of GR unaffected.
That is, GR can be extended all the way to the Planck scale in its usual
formulation of $G_{\mu\nu}+\Lambda g_{\mu\nu}=8\pi GT_{\mu\nu}$, for the
Einstein tensor $G_{\mu\nu}$, the cosmological constant $\Lambda$, the
metric tensor $g_{\mu\nu}$ and the energy-momentum tensor $T_{\mu\nu}$.

This is because the evolution of $n_1$ due to Eq.~(17) finds a natural place
in $\Lambda$ (sections 5.1 and 6.2), leaving $G_{\mu\nu}$ unaffected. While
the statistical properties (sections 6.3 and 6.5) of $T_{\mu\nu}$ in its
standard form are completely specified by topological dynamics.
Therefore, the beautiful properties of GR are completely preserved under
the topological extension further explored below.

\subsubsection{The Initial State and Superposition}

The natural initial condition for Eq.~(17) is a single $T^3$ as the start of
a multiply connected 4-space at $m=0$. On this $T^3$ quantum fields can live
with matter degrees of freedom and therefore the possibility to form the
first BH(s) through GR.
This way, quantum superposition in the form of distinct paths
on a three-torus in 4-space is what forms the initial cause for our universe.

Multiplied three-tori and induced BHs, in every time slice $m$, carry
no spatial information. When created, the connected sum assures that all
3-primes within a time slice are linked through junctions homeomorphic to
$S^3$. Thus assuring that there are no island universes, but only a single
path connected 4-space.

\subsection{A Topological Extension of Mach's Principle}

Before solving Eq.~(17) under certain circumstances, there is an immediate
result that can be read off in the context of Mach's principle.
The quantum foam stability problem that plaques GR on the Planck scale, as
explored by Wheeler, may find a natural resolution in the quantum space-time
equation of motion above. If one considers the limit where

\begin{equation}
\Delta n_1=\Delta M=\Delta F\approx 0,
\end{equation}
so a static or very slowly evolving population of handles with little
classical BH formation and merging due to matter degrees of freedom, then

\begin{equation}
2n_1\approx \Delta E.
\end{equation}

The quantum foam instability problem is solved {\it if} $n_1$ in Eq.~(19)
pertains to long-lived BHs. I.e., if there is an ensemble of BHs that lives for
discrete times $mt_P$ at least comparable to the current age of the universe,
then this population of $n_1$ long-lived BHs must cause the evaporation of
twice that many induced BHs every Planck time.
Afterall, for $n_1(m+1) \approx n_1(m)$ the
evaporation term $\Delta E$ must act on not more than
the Planck time for consistency.

The natural mass for these associated mini BHs is the Planck mass, if they are
to evaporate in a Planck time.
This while the evaporation time of a BH scales as its mass to the third power.
So a BH of more than $2\times 10^{15}$ g would live as long as 10 billion
years. Hence, an ensemble of macroscopic BHs provides a naturally stable
quantum foam of Planckian mini BHs because of the MP. This leads to the
quantum foam that Wheeler envisioned, but one that is constrained by the
longevity of macroscopic BHs.

Einstein's interpretation of Mach's principle states that the global
distribution of matter somehow determines the (changes in) local geometry,
and vice versa. Wheeler's local quantum foam of mini BHs, with its multiply
connected topology, is an expression of GR. Therefore, following Einstein's
ideas for geometry with the MP for topology, matter that is globally observed
to be in the form of macroscopic BHs somehow determines the (changes in)
Planck scale topology, and vice versa.
One may thus go one step further and use the MP to derive the topological
extension of Mach's principle (EMP). It is formulated as follows.
\medskip

{\it The global topology and geometry of the universe determine the changes
in Planck scale topology and the local motions of matter, and vice versa.}

\medskip
This EMP is in essence the answer to the following question:
How does Nature know what quantum fluctuations in geometry to provide for,
in every Planckian volume? The answer suggested here, is that the collective
action of all long-lived BHs provides this. Although strange, the EMP is in
the spirit of Mach's and Einstein's ideas, with some help of the MP.

The number\footnote{Not the number density.}
of macroscopic BHs in the entire universe, so in all of 3-space for some
time slice through 4-space, is therefore the global quantity that determines
the local occurrence of mini BHs in Wheeler's quantum foam.
Because of homeomorphic invariance, there is no
specific volume to which induced mini BHs couple. However, despite
being global principles, the MP and the EMP together do constrain this volume.

To speak of a Machian universe in the extended topological sense, there should
be a stable global topology. That is, the inducing population of macroscopic
BHs should have a longevity $m't_P$ that exceeds the age $mt_P$ of the
universe at that time (the existence of a global time coordinate is addressed
in section 4 below).
Furthermore, the MP dictates that Eq.~(19) already holds for
a single long-lived BH and that any local observer of the quantum foam can
determine the (growing) number of such macroscopic BHs.

This renders the first occurrence of a long-lived BH a locally and globally
observable quantity on $L(T^3)^+(m)$ in 4-space. I.e., paths that enter such
a BH connect to a topologically distinguishable future $m'>m$.
So, if $L_f$ is the size of the universe when the first BH forms that exists
for longer than the contemporary age of the universe, then $L_f$ is frozen in
and the quantum foam stabilizes.
That is, even though topological in origin, $L_f$ in Planck lengths is above
all a geometric number that quantifies the first embedding volume $L_f^3$ of
our universe in 4-space.
The scale $L_f$ will be re-addressed in the context of dark energy in section
5 below.

\subsection{Some Further Solutions: Exponential Self-Multiplication and Decay}

Besides the Machian solution above, Eq.~(17) can be also be solved
straightforwardly under other approximations.
For example, if $\Delta F=\Delta E=\Delta M\approx 0$ then

\begin{equation}
n_i(m+1)-n_i(m)\approx \mu_i n_i(m),
\end{equation}
with the approximate exponential solutions

\begin{equation}
n_0(m) \approx 7^m
\end{equation}
and

\begin{equation}
n_1(m \ge x) \approx 3^{m-x},
\end{equation}
for initial conditions of the universe at $m=0$ that involve a single
$T^3$ and with a first $S^1\times S^2$ formed by matter degrees of freedom
after a short time $xt_P$ ($x<10$).
Hence, quantum space-time can experience a self-multiplication phase that
very quickly builds up a lattice of three-tori with an (unstable) foam of
mini BHs attached to it. Such a state for the quantum foam will re-addressed
in the context of inflation in section 6 below.

For the three-tori, which are always confined to the Planck scale contrary
to the BHs, there are no further solutions unless one allows for reversals of
the arrow of time. For the handles, there is also the situation where
$\Delta F <<\Delta E$ and $\Delta M <<\Delta E$ over extended ($m>>1$) periods.
In this limit one has that

\begin{equation}
n_1(m+1)-n_1(m)\approx \mu_1 n_1(m) -\Delta E,
\end{equation}
and for $-\Delta E + 2n_1(m) \equiv -D n_1(m)< 0$, with $D<<1$, that

\begin{equation}
n_1(m+1) - n_1(m)\approx -Dn_1(m),
\end{equation}
if the radiation background in the universe is low enough to allow all induced
mini BHs on the right hand side of Eq.~(24) to evaporate in a Planck time or
so.

This yields the approximate decay solution for times $m>m_1$

\begin{equation}
n_1(m>m_1) \approx N_1(m_1) e^{-mt_P/t_e},
\end{equation}
if there is a longer-lived population of BHs, at time $m_1$ and $N_1$ in
number, that evaporates on a time scale $t_e = D^{-1} t_p$ with $t_e/t_P<m_1$.
I.e., one can think of a population of $N_1$ primordial BHs, with masses much
larger than the Planck mass but much smaller than $\sim 10^{15}$ g, in a
cooled/cooling down universe.

\section{The Global Nature of Time}

The equation of motion derived above is discrete in time, but with continuous
time intervals, because the MP is
based on the concept of counting. One may feel uncomfortable with this
approach to the nature of time.
However, it is possible to use the $T^3$ solution in Eq.~(21), and the
identifications on $L(T^3)$, to provide a more solid foundation for time as
a global topological notion as well as an approximate geometric variable.

\subsection{Counting Three-Tori}

Unlike handles, three-tori always increase in number monotonically
as $n_0=7^m$. Turning this around, it is the homeomorphically
invariant number $n_0$ that is fundamental and globally known according to
the MP. Hence, a sense of topological time is naturally defined as the
quantity $m\ =\  ^7log(n_0)$, or $mt_P$ in physical units. I.e., the MP
{\it endows} continuous 4-space with a countable 3-space structure.

It is not the case that every observer goes about the task of counting
three-tori to experience some sense of time. Afterall, time is merely the
parameter to denote the local changes that the universe undergoes from one's
own perspective. This while the number $n_0$ is currently
$n_0(m_0)\sim 7^{610}$, for an of age $m_0$ of the universe of about
14 billion years (or $\sim 10^{61}t_P$).

Rather than counting the number $n_0(m_0)$, one concludes that as many as
$n_0(m_0)$ degrees of freedom are required to exhaust the state space
capacity of the lattice of three-tori, and to expose its current fine-grained
structure. With $\sim 10^{80}$ particles in the universe, this is not likely
to occur and even a multiverse could be supported by $L(T^3)^+(m_0)$.

\subsection{The Appearance of Relativity}

The global notion of time explored above, based on counting, may seem at
odds with the usual local relativistic nature of time. However, one actually
complements the other. The number of three-tori, like the number of birds in
the sky, is a Lorentz invariant. Information on the number of birds can only
travel at the speed of light, so has a local effect. At the Planck scale, as
mentioned above already, the lattice of three-tori enjoys homeomorphisms.
These allow any two quantum space-time points to be brought into close
proximity, facilitating information transfer without exceeding the speed of
light. On scales much larger than $l_P$, the diffeomorphism invariance of
$S^3$ should dominate. The latter limit can be established as follows.

The MP preserves the feature of every $T^3$ that it is the boundary of a
Lorentz 4-manifold. Subsequently, $L(T^3)(m)$ yields a Lorentzian signature
$-+++$ for the 4-space that it bounds, through every Planckian $T^3$
volume in a time slice through 4-space. Because $S^1\times S^2$ has a
zero Euler characteristic as well, the lattice $L(T^3)^+(m)$ also
guarantees a quantum space-time with the local causal properties envisioned
by Minkowski. One thus finds that quantum superposition and the MP lead to
a lattice of three-tori that impose a metric signature appropriate for
relativity.

Similar to the connection between $L(T^3)$ and $T_n$ for topological dynamics
and string theory, there is a relation between topological time $m$ and
geometric time $t$. The handles are ignored in what follows, so $L(T^3)(m)$
instead of $L(T^3)^+(m)$ is used, because on large scales macroscopic BHs are
local perturbations only.

Using the homeomorphic equivalence $T^3\oplus S^3\sim T^3$, one can
write $L(T^3)$ as

\begin{equation}
\oplus^i [T^3_i \oplus S^3_i],
\end{equation}
and consider the limit in which the scale $R$ of $S^3$ obeys $R>>l_P$.
Observers that measure with modest frequencies, so $<<1/t_P$, are
not able to detect the rapidly varying phases of the $S^1$ loops in a $T^3$.
That is, any global number information on three-tori (as well as handles
and their longevity) is carried by 4-space topological identifications.
The latter become sampled at very poor temporal resolution,
with a duty cycle $\delta_t = 1/m$. In a universe with a speed of light
that is empirically found to be constant, one has $R(m)=ml_P$ in physical
units and can write, for $i=1$ to $i=m^3$, that

\begin{equation}
L(T^3) \approx [\oplus^i T^3_i]/m^3 \oplus S^3|R,
\end{equation}
in which $S^3|R$ is a three-sphere with geometric scale $R$. The $m^3$
normalization of the $T^3$ sub-lattice $|L(T^3)|\equiv \oplus^i T^3_i$
signifies its sampling weight (analogous to $T_n$ in section 2.3.2 above),
for three independent
directions that are each sampled with a $1/m$ topological time duty cycle.

As in section 2.3.2, the $1/m^3$ weight augments the connected sum with a
proper relative count of the components that it acts on, independent of
homeomorphic invariance. In Eq.~(27), the relative
multiplicity of components is thus $1/m^3$ $|L(T^3)|$ to 1 $S^3|R$. The weight
of the sub-lattice $|L(T^3)|$ signifies that a very lethargic observer
on $S^3|R$ can only count one of the $m^3$
identifications on $|L(T^3)|$ during a topological time $m>>1$.

Consequently, one can make the step to the large scale geometry of space-time
through

\begin{equation}
L(T^3) \approx T^3|[R,\delta_t^{-1}] \oplus S^3|R,
\end{equation}
in which $T^3|[R,\delta_t^{-1}]$ is a three-torus with geometric scale
$R=ml_P$ that requires a geometric time $t=t_P/\delta_t$ to causally
traverse its internal solid cube 3-space. Time enters the discussion
relativistically here because for
$R>>l_P$ (so $m>>1$) the propagation of information should proceed causally,
at the speed of light, as if the universe is simply connected.
For any universe with a size law $R(t)$ that is monotonically increasing in
geometric time $t$, the limit $R\rightarrow\infty$ is well defined for any
observer. Locally, so for $d<<R$ but $d$ and $R$ both very large, one has

\begin{equation}
L(T^3)\approx S^3|d
\end{equation}
since $T^3|[R,\delta_t^{-1}]$ appears simply connected on scale $d$.

Globally, so on scale $R$ for any finite $t$,

\begin{equation}
L(T^3)\approx T^3|[R,\delta_t^{-1}],
\end{equation}
since $S^3$ is simply connected. In physical units,
the topological time $mt_P=t_P/\delta_t$ becomes the geometric time $t$ for
$m\rightarrow\infty$, so $\delta_t\rightarrow 0$.
This occurs because a three-torus is always the
boundary of a Lorentz 4-manifold (it has a zero Euler characteristic),
while $T^3|[R,\delta_t^{-1}]$ limits to a locally flat and locally simply
connected geometry for large $m$.

The local differential structure on this Lorentz 4-manifold is
stable, and provided by GR, only for time intervals $\Delta m$ that obey
$2\Delta m<m$. Under the latter condition two three-tori of spatial size $R$
and time size $\Delta m$ are insufficient to cover the full time interval $m$
through topological identifications across $\Delta m$. Subsequently, at
least one more three-torus of spatial size $R$ and time size $t_P$ is
necessary. The latter Planckian time slice can be placed anywhere in
4-space and allows one to
represent every discrete time $m'\in (1,m)$, on spatial scale $R$, by
differencing relative to $\Delta m$.

So the diffeomorphic invariance of shape and a geometric notion of time
result from the limit of large
$R(m)$ and small $\delta_t = 1/m$, but only for observers that enjoy
$2\Delta m<m$. For $2\Delta m\ge m$ an observer would see the
coarse-grained nature of time through the topological identification of two
different time-like slices over an interval $\Delta m >>1$ in $L(T^3)(m)$.
See also sections 6.4 and 7.1 for further consequences of this topological
feature of time.

\section{Dark Energy}

In order to test the idea of topological dynamics in a definitive manner,
pertinent predictions should be made. To this effect, dark energy is now
studied. I.e., the expansion of the universe appears to be accelerating, as
indicated by the magnitude-redshift relation of type Ia
supernovae, over the past $\sim 7$ billion years.$^{17,18,19}$

This negative gravity phenomenon constitutes one
of the biggest mysteries in cosmology. Dark energy appears to be fundamentally
connected to the dynamics of quantum space-time.$^{12}$
The simplest expression of
dark energy is in the form of a positive cosmological constant term on the
left hand (so geometric) side of the Einstein equation. Nowadays, dark
energy is often included on the right hand (so energy-momentum) side of the
Einstein equation, as a fluid that possesses the property of negative pressure.
The origin of a cosmological constant (or any other) term is not specified by
GR, leaving one with an incomplete description of space-time dynamics.

\subsection{The Globally Embedded Quantum Foam of Wheeler}

The handle multiplicity evolution discussed above leads to the unambiguous
prediction that macroscopic BHs stabilize the quantum foam of Wheeler.
The induced mini BHs of Eq.~(19) require an increase in 4-volume for their
embedding. This immediately yields the correct sign for a cosmological
``constant'' term on the metric side of the Einstein equation.

That is, this form of negative gravity has a topological origin. Any observer
on $L(T^3)^+(m)$ sees the formation of an induced mini BH as the spontaneous
bulging out of a locally $S^3$ piece of 3-space into a $S^1\times S^2$ part
of {\it future} 4-space, under the act of BH observation.
Hence, the MP induced mini BHs pull 3-space into 4-space and the proper sign
for a dark energy term is obtained purely topologically.
In this, and as mentioned before,
induced mini BHs appear randomly throughout a time slice, so at at
any spatial position, changing only the integer properties of the wave
function of the universe $\psi$.

Furthermore, the number of macroscopic BHs is a function of time and may
vary from one time slice of thickness $t_P$ to the next.
Dark energy, as an evolutionary expression of vacuum energy, is therefore
associated in this work with the mini BHs that are induced by long-lived BHs. 

To be more quantitative about this, recall that the MP and the EMP lead to a
specific scale $L_f$ equal to the size of the universe when the first BH was
formed that lives longer than the age of the universe at that time. With
$N_{\rm BH}(m)$ the number of macroscopic BHs as a function of topological
time, one then finds from Eq.~(19) a dark energy density of

\begin{equation}
\Lambda (m) = 2\frac{N_{\rm BH}(m)\ m_P}{L_f^3}.
\end{equation}

Note here that the topological time $m$ manifestly expresses that no
violation in causality occurs as information on the total number of
macroscopic BHs is distributed by the irreducible collapse of $\psi$ across
time-like slices of 4-space on the Planck scale. Time is still relative of
course and one may freely choose any foliation for the time slices $m$ since
the number of BHs is a generally covariant quantity. In cosmology it is
common to use the redshift $z$ for this, where $1+z$ measures the
change in scale factor of the universe. So $N_{\rm BH}(z)$ will be used
in the remainder of this section.

Furthermore, the evaporation of mini BHs typically occurs within about a
Planck time of the ambient environment is cold. Hence, the dark energy
density of the stabilized quantum foam is a true 4-space vacuum
energy, with its place on the left hand side of the Einstein equation.
Indeed, mini BH hole decay does not automatically lead to a particle
evaporation signature that can be observed by a macroscopic long-lived
observer (somewhere in the {\it entire} universe).
Rather, it requires a detailed solution of GR to determine
which induced mini BHs can live for longer than a time interval $(m+)$
(see further comments in section 6).

Eq.~(31) shows that, for $N_{\rm BH}(z)$ independent of redshift, any local
observer concludes that the dark energy density $\Lambda$ in g cm$^{-3}$
stays constant as the universe expands. I.e., a true cosmological constant
is allowed for this topological form of dark energy, but this depends on the
BH number evolution of the universe.

\subsection{Dark Energy and the BH Formation History of the Universe}

As is evident from above, the dark energy density of our universe is linearly
proportional to the total number of macroscopic BHs in three-space at any
redshift labelling a time slice through 4-space.
The bulk of the stars, and thus stellar BHs, in the universe appears to be
present as early as $z=1$, when a $(1+z)^4$ decline ensues in the cosmic star
formation rate$^{27}$ after a peak at $z\sim 2$. Roughly 1\% of all formed
stars are so-called high-mass ($>8 M_\odot$) stars.

BHs are the
remnnant of such high-mass stars. One therefore expects to find an effectively
constant $\Lambda$ for $z\le 1$ because the bulk of all stars has been
formed by then. Conversely, given the rapidity with which
massive stars/BHs are produced during $z=1-3$, one expects a strong decrease in
the number of BHs, and thus in $\Lambda $, from $z=1$ to $z=3$.

These qualitative expectations can be quantified as follows.
In Ref.~27, their figure 7, the comoving type II supernovae (SNe) rate
density is derived from the star formation history of the universe.
In this, it is assumed that macroscopic (so stellar mass) BHs are
produced by type II SNe with some constant efficiency (roughly 10\%).
Furthermore, the initial mass function of stars is taken to follow the
(universal) Salpeter shape.

For a local observer today, it is plausible to consider $\Lambda (z)$
normalized to its current value $\Lambda (0)$, where the latter
corresponds to the present number of BHs in the entire universe.
The ratio $\Lambda (z)/\Lambda (0)$ can be expressed in
terms of $dN_{\rm BH}/dz$, using the data in Ref.~27 and for a comoving
volume$^{28}$, as

\begin{equation}
\Lambda (z)/\Lambda (0) = \frac{\int_y^z dt\ dN_{\rm BH}/dt}{\int_y^0 dt\ dN_{\rm BH}/dt} \approx (1+z)^{-0.36}\ {\rm for}\ z<1.
\end{equation}

This is independent of a constant BH formation efficiency. Here,
a flat WMAP9 cosmology is adopted, with macroscopic BH formation starting
at some large redshift $y>30$ but at a very modest pace, and the
usual cosmic time interval $dt(z)$ is used.

Although a redshift coordinate is adopted for different time slices through
4-space, there is no pertinent dependence
on coordinate system because of the normalization in Eq.~(32). Any observer,
whatever its coordinate system, finds the dimensionless ratio of two numbers
to be generally covariant. It is assumed here that any observer can derive the
temporal change in the total number of BHs, as three-space evolves, because
he lives in a representative comoving volume that undergoes the same relative
evolution in BH formation as the entire universe.

One finds numerically that $\Lambda (z)/\Lambda (0)$ roughly scales
as $(1+z)^{-0.36}$ for $z<1$. I.e., $\Lambda (z)$ changes by less than
$\sim 30$\% for an epoch more recently than a redshift of unity.
The value of the equation of state parameter $w=p/\rho$, for pressure $p$,
density $\rho$ and constant $w$, is estimated through WMAP9 data to be
$w= -1.08\pm 6\%$ for $z<1$.$^{29}$ This while recent PLANCK results yield
$w=1.13\pm 10\%$.$^{30}$

These observational data are consistent with Eq.~(32). This is because a
constant value of $w= -1.1$ is equivalent to a change in
$\Lambda (z)/\Lambda (0)$ of about 30\%, for $z<1$. The latter follows from
the conservation equation

\begin{equation}
\frac{d\rho /dt}{\rho}=-3H(1+p/\rho )=-3H[1+w(z)],
\end{equation}
with the Hubble parameter squared $H^2=8\pi\rho /3$.

If one expands $1+w$ as $1+w\approx \delta w$, around $w\approx -1$, then
the relative change in $\rho$ over a local Hubble time is about
$(H\rho )^{-1}d\rho /dt=-3\delta w$, for constant $\delta w$.
Conversely, because the majority of the stellar mass BHs appears not to have
been formed before $z=1-2$, one expects to have
$w < -1$ for $z>1-2$. Figure 1 shows a graphic representation of Eq.~(32).
The shading in Fig.~(1) represents the observational uncertainty in deriving
the type II SNe rate from the cosmic star formation history, when a Salpeter
initial mass function is adopted.$^{27,28}$

It is possible that many, comparable to the stellar mass BH count, primordial
BHs exist$^{15,31}$ that start to evaporate only for $z<3$. If so, then values
of $w$ larger than $-1$ are possible during that cosmic epoch because the dark
energy density diminishes as $N_{\rm BH}(z)$ goes down with decreasing
redshift $z$.
Such primordial black holes can also be a result of the exponential
self-multiplication discussed above. They only survive until today if these
primordial BHs are more massive than $\sim 2\times 10^{15}$ g.

\begin{figure}[!htb]
\begin{center}
\includegraphics[angle=0,width=14cm]{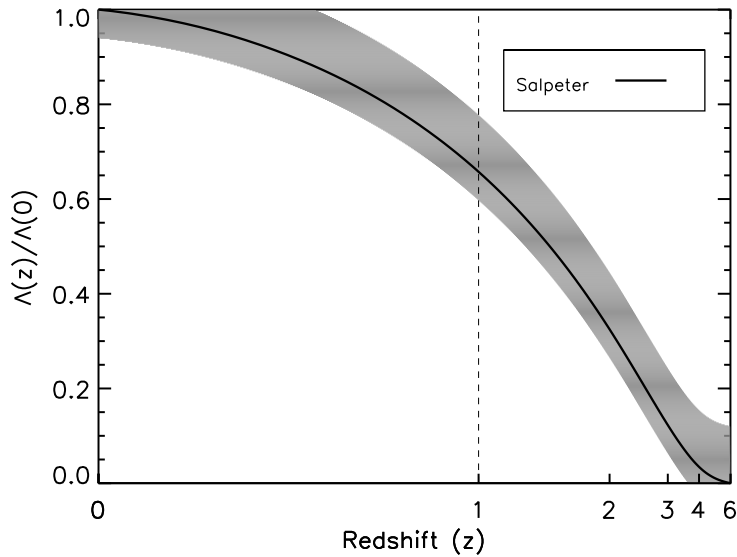}
\caption{Normalized redshift evolution of the dark energy density, denoted by $\Lambda (z)$ and normalized to its present-day value $\Lambda (0)$.
The shading in the figure represents the observational uncertainty in deriving
the type II SNe rate, when a universal Salpeter initial mass function is
adopted.}
\label{fig:DE}
\end{center}
\end{figure}

In Ref.~20 it is pointed out that, although $-w$ is close to unity, there
is room for evolution during $z>0.5-1$. This is because the SNe Ia data only
weakly constrain dark energy for such early times (there are still only a
modest number of them detected).
Furthermore, recent WMAP7 work$^{32}$ provides limits on the properties of
time dependent dark energy parameterized by

\begin{equation}
w(a)=w_0+w_a(1-a),
\end{equation}
where $a$ is the scale factor of the universe ($a=\frac{1}{1+z}$), and
$w_0=-0.93\pm 0.12$ with $w_a=-0.38^{+0.66}_{-0.65}$,
for a flat cosmology.

The 13\% uncertainty in $w_0$ is easily accommodated
by the shading in Fig.~(1), while the fiducial range of $w_a$ allows for a
factor of 5 change in $\Lambda$ over $z=1-3$.
Recent results by Ref.~33 yield $w_0=-0.905\pm 0.196$ and
$w_a=-0.984^{+1.094}_{-1.097}$, which are again consistent with Fig.~(1).

In any case, it is apparent that a modest improvement in future measurement
precision of an effective $w$ around $z\sim 1$ could confirm or rule out
the topological
dynamics model for dark energy advocated here. Furthermore, the one-to-one
correspondence between the number of macroscopic BHs and the dark energy
density is very much open to astrophysical detection, given the accuracy
with which the cosmic star formation history can be probed.$^{27}$
If verified, then this one-to-one link constitutes the proverbial
{\it smoking gun} in support of the validity of the MP and the EMP.

\subsection{The Spatial Scale and Timing of Dark Energy}

The current value of the dark energy density $\Lambda$ is
$\sim 10^{-29}$ g cm$^{-3}$.$^{29}$ Because $\Lambda =2N_{\rm BH}m_P/L_f^3$,
the relation between the dark energy density and BH number can be used to
compute $L_f$.

The present number of macroscopic (mostly stellar mass) BHs in the entire
universe is approximately
$N_{\rm BH}\sim 10^{19}$.\footnote{There are about $10^{11}$ galaxies with
each about $10^{11}$ stars. Roughly 1 in $10^3$ stars ends its life as a
BH remnant.}
From the present-day value of the dark energy
density it then follows that $L_f\approx 2\times 10^{14}$ cm, about the
size of the Solar system. Hence, the physical scale of the dark energy density,
not specified in the topological dynamics studied here, is determined by
empirical data. This renders the form of quantum gravity presented here fully
fixed.

A number of other concerns regarding dark energy find a natural resolution
when macroscopic BHs induce mini BHs. No modifications to matter degrees of
freedom or additional (scalar) fields, like
in quintessence as well as k-essence and Chaplygin gas approaches, are
required or need to be justified in order to introduce dark
energy.$^{34,35}$ In fact, because the star formation history
of the universe sets the appearance of present-day dark energy, it is
ultimately the Einstein equation itself that determines the evolution of
$\Lambda $ (or some effective $w$).

Furthermore, since
BHs are stellar remnants, it should come as no surprise that we witness dark
energy in our time. Today is part of an epoch during which the bulk of the
stars in the universe has been formed. This while the most massive of those
stars have already exploded as SNe to leave behind BHs.

The change in the magnitude of the current dark energy density is set by the
number of macroscopic BHs. With the bulk of all stars formed, so with much of
the baryonic matter in galaxies turned into stars, this number cannot increase
much further. We thus find ourselves, orbiting a typical star, close to the
maximum dark energy density in time. This is for some value of $L_f$.
The fact that the dark energy density is roughly comparable to the total
matter density, in units of the critical density, then seems an odd
coincidence.

However, a larger embedding volume $L_f^3$ for the universe diminishes
$\Lambda$ as much as it increases the total mass in the universe available
to make BHs. So $\Lambda$ is mostly affected by the mean mass of a BH as well
as the fraction of baryonic matter that is converted into stars and
subsequently into BHs. For the present universe these numbers are about a few
Solar masses and 0.1\%, respectively. Interestingly, stellar physics forbids
much lower mass BHs because electron or neutron degeneracy pressure provides
stability. A much larger BH mass or fraction would require a very exotic (and
not observed) stellar initial mass function.

Hence, the value of $\Lambda\sim 10^{-29}$ g cm$^{-3}$ is a robust number,
within 1-2 orders of magnitude, for an epoch after the peak in the cosmic star
formation rate. This assumes that stellar remnants are the main source of
long-lived BHs. The mean cosmic density, with its steep $(1+z)^3$ dependence,
is currently about a few $\times 10^{-30}$ g cm$^{-3}$ and increases by
orders of magnitude in the past.
This while the universe has spend most of its cosmic life after the peak in
the star formation history. This peak occurs for $z=1-2$, at least 9 billion
years ago, but also billions of years after the big bang.

The densities of dark energy and matter are therefore roughly and naturally
comparable for $z\sim 1-2$ if the bulk of the stars/BHs are in place.
Also, we are likely to find ourselves after the peak in
the cosmic star formation rate. The latter favors, through cosmic expansion,
the dark energy density by a modest factor relative to the total matter
density.

Hence, both the timing and magnitude of $\Lambda$ are as one would expect.
An anthropic principle does not appear to be needed in any strong sense.
Had the Earth been formed already a few billion years after the big bang,
before the peak in the cosmic star formation history, then $N_{\rm BH}$ had
simply not yet reached its maximum. Still, dark energy would not be completely
absent either.

\section{Inflation}

The inflationary paradigm has shaped modern-day cosmology and knows many
different incarnations.$^{36,37,38,39,40}$ To further constrain topological
dynamics, this section aims to link the solutions of the equation of motion
above to a possible inflationary phase of the universe.

\subsection{Number of e-Foldings and Present-Day Dark Energy}

In the early universe, before the first
long-lived BH stabilizes the quantum foam, Eq.~(22) presents an exponential
solution of the equation of motion for the handle multiplicity. Just like
Wheeler's stabilized quantum foam can be associated with dark energy, so can
the young and unstable quantum foam be used to drive inflation.

For inflation to continue, so for Eq.~(22) to pertain, it is necessary that
the handle multiplicity is maintained. I.e., that the ambient matter and
radiation density are high enough to suppress induced mini BH evaporation.
The latter
requires a detailed exploration of topologically extended GR.
Such a study is postponed to a future paper. Nevertheless,
if topological dynamics has any merit, then it is possible to compute the
number of e-foldings during inflation from $L_f$, as follows.

The inflationary expansion of the
universe is one of (infinitely) many homeomorphisms that space-time enjoys.
Eq.~(17) is invariant under homeomorphisms because of its topological nature
(the MP). However, the form of Eq.~(17) should also lead to the correct
geometric limit because the longevity of macroscopic black holes yields
the notion of a specific size scale $L_f$ (the EMP).
Hence, one should seek a geometrization of Eq.~(17).

At this point, the topologically void sub-equation of motion for $S^3$
actually becomes useful. Recall that

\begin{equation}
\Delta n_2 S^3 = \delta_2 n_2 S^3 = 0,
\end{equation}
and consider topological time in the limit $m>>1$ and $\Delta m>>1$, written
as $\tau$ for $m$ and $d\tau$ for $\Delta m<m$, during inflation. The value
of $\Delta m$ need not be small when using the variable $\tau$. In physical
units $\tau =mt_P>>t_P$, with an associated physical scale
$R(\tau )=mt_P/c=ml_P$ (the same as $R(m)$). It is
important to distinguish $\tau$ from geometric time $t$ here, because
$2\Delta m<m$ does not hold for all $d\tau$. I.e., a priori, one considers
all time intervals in the universe's history.

One can now turn to the geometrization of Eq.~(35).
Although yielding a zero in terms of counting loops, geometrically one can
represent the three-sphere by the relative size of the universe $S(\tau )$
for $n_2(m)\rightarrow S(\tau )$. That is,
$S(\tau )$ measures the ratio (a dimensionless number) of the size of
the universe with respect to
its pre-inflationary state. Afterall, any notion of geometry on $S^3$,
irrespective of a metric, should allow one to count Planck lengths if time
is intrinsically discrete and if the speed of light $c$ is empirically
found to be a universal constant.

The action of $\Delta$ is a first order time difference, which can now be
written as $dS/d\tau$, for changes that are not necessarily small.
The action of the counting operator $\delta_2$ only has meaning in
conjunction with $n_2(m)\rightarrow S(\tau )$ and must yield
$\delta_2S\rightarrow 0$ when one takes the limit $S\rightarrow\infty$.
Therefore, $\delta_2$ must be representable as a
function of $S$ alone and be a decreasing function thereof. Inflation
is a homeomorphism that drives the universe to a flat state. The
geometrization of $\delta_2$ thus represents vanishingly small curvature.
Hence, $\delta_2\rightarrow 1/S^2$ and one finds that Eq.~(35) turns into

\begin{equation}
\frac{dS}{d\tau} = \frac{1}{S}.
\end{equation}

The solution $S(\tau )=(2\tau )^{1/2}$ is easily found.
Because $\tau_f=L_f/l_P$ when $L_f$ becomes frozen in, one obtains
$S_f=2^{1/2}(L_f/l_P)^{1/2}$. Note that $L_f\approx 1.3\times 10^{47}$
in units of $l_P$, so that $S_f\approx 5.0\times 10^{23}$.
One therefore finds that there is a natural {\it correspondence}. The
pre-inflationary scaling law $R(t)\propto t^{1/2}$, for a local observer
experiencing geometric time $t$, has the same form as the inflationary period
scaling law $S(\tau )\propto \tau^{1/2}$, expressed in topological time $\tau$.
This correspondence between $t$ and $\tau$ shows that a topologically driven
inflationary period leads to a specific geometric embedding in 4-space for
our universe.

It is good to realize here what the physical meaning of $\tau_f$ and $L_f$ is.
The MP leads to the $L(T^3)^+(m)$ structure of quantum space-time, which
possesses a Lorentzian signature (see section 4.2). The large $m$ topological
time $\tau$ then indicates a topological size $ml_P>>l_P$ on the lattice.
Only when the first long-lived BH forms, can one speak of
our universe as an entity of size $L_f$ that exists on $L(T^3)^+$ with a
distinct future $m'>m_f=L_f/l_P$.

Since $L_f\propto N_{\rm BH}^{1/3}$ it follows that the increase in this
characteristic size obeys $S_f\propto N_{\rm BH}^{1/6}$. This is a weak
dependence. The number of e-foldings n scales as
${\rm n}\propto log N_{\rm BH}^{1/6}$ for $S_f\equiv e^{\rm n}$.
The current value of $N_{\rm BH}\sim 10^{19}$ for stellar mass BHs in the
observable universe yields ${\rm n}\approx 55$.

The stellar initial mass function in active galaxies or
the early universe may have been top-heavy.$^{41}$ Such a non-Salpeter shape
favors high mass stars and can boost $N_{\rm BH}$ by easily an order of
magnitude. Still, this will have a limited effect on n.
Of course, the observable universe might be a lot smaller than the entire
universe. Nevertheless, even a boost in $N_{\rm BH}$ by a factor of $10^6$
leads to an increase in n to $\approx 57$.
In any case, these values are within the fiducial Planck range$^{42}$ of 50-60.

\subsection{The Onset of Inflation}

A form of inflation that is driven by mini BHs suggests a start that picks
up right after the big bang. However, this is not the case.
First, one needs an ensemble of small BHs that are long-lived enough.
In the sense that the mean lifetime of the ensemble is at least much larger
than a Planck time.
I.e., so that the handle multiplicity can at least be maintained, as already
mentioned above. These BHs are referred to as longer-lived mini BHs.

The dark energy density is linearly proportional to the number of such
longer-lived mini BHs, denoted as $N_{\rm BHll}$.
Because $L_f$ is not yet frozen in at some time $mt_P<m_ft_P$, it is not
obvious how to determine the specific volume $l^3$ for the dark energy density

\begin{equation}
\Lambda (m)=2N_{\rm BHll}\ m_P/l^3.
\end{equation}

But, this characteristic length $l$ must be approximately set by the
Schwarzschild radius $R_{\rm Sll}$ of the longest living mini BH,
$l=R_{\rm Sll}$.
The latter radius is the proper local length scale in topological dynamics.
The number of Planck times $1<k<m$ in a BH's existence is expressed through
the number of three-tori.
The longest living BH then sets $l$ because its life span
comprises many more three-tori and their connections on $L(T^3)^+(m)$ than do
mini BHs with life times $k'<k$.

Because $n_0(k)/n_0(k')=7^{k-k'}$, the longest living BH dominates the
$L(T^3)^+$ state space in terms of the information (entropy) required to
describe it in 4-space. The longest living BH thus
defines the most likely entropic state that the universe finds itself in.
The precise number and characteristic mass of longer-lived mini BHs requires
a full exploration of Eq.~(17) within GR.\footnote{As argued in Ref.~12, a
BH can grow by induced mini BHs even when no matter passes its event horizon.
This occurs at a maximum rate of
$\sim 1\ (N_{\rm BH}/10^{19}) \ M_\odot {\rm year}^{-1}$
for BHs of $10^9 M_\odot$ that have a Schwarzschild radius equal to $L_f$
(set in turn by the first long-lived BH), and with a
linear mass dependence below $10^9 M_\odot$. The bulk of the mass of a BH
like the one in the Milky Way center can be acquired this way.}
Still, one can try to make an estimate of the typical longer-lived mini BH
mass to obtain a grip on the timing of inflation.

Grand unified theory has a characteristic energy scale of $E_c\sim 10^{16}$
GeV, or about $\epsilon\sim 10^{-3}$ of the Planck energy. During this
epoch, the universe has substantially cooled down in energy scale $E$ relative
to the Planck energy, while primordial BH production may naturally occur
during the associated phase transition. The thermodynamic temperature
$T_{\rm BH}$ of a BH scales with its mass $m_{\rm BH}$ as
$T_{\rm BH}\propto 1/m_{\rm BH}$, and thus its typical life time $t_e$ as
$t_e\propto m_{\rm BH}^3$.
It is therefore reasonable to estimate that one has longer-lived mini BHs with
$m_{\rm BHll}\sim m_P/\epsilon$ that live for about
$t_e\sim t_P/\epsilon^3\sim 10^9t_P$.

Consequently, one has a duration of at most $t_e$ to build up an ensemble of
longer-lived mini BHs that can act as a base for inflation.
Conversely, the size scale $R$ of the universe grows as $R\propto t^{1/2}$,
so $t\propto R^2$, prior to inflation. It then takes a cool down period $t_c$
of about $t_c\sim t_P/\epsilon^2\sim 10^6t_P$ to reach the grand unification
energy scale, since $E\propto 1/R$.
In all, one expects inflation to commence after $t_c$ and before $t_e$.
This yields a fiducial range of about $10^6-10^9 t_P$. Although broad, the
square root mean value is in line with general expectations.$^{7}$

\subsection{The Persistence of Homogeneity}

There is also the issue of the initial conditions for inflation and how
likely these are.$^{43}$ In units of $t_P$ a long time passes before mini BH
driven inflation commenses. Over this time, some region of the vacuum
that is to become inflated need not be in a state of homogeneity. If it
has to be, then this renders the initial conditions quite unlikely.

However, the combination of the MP and the EMP form the topological heart of
mini BH driven inflation. Together they dictate that every local Planckian
volume responds, though the induction of Planckian mini BHs, in the same
manner to the global ensemble of longer-lived mini BHs.
This joint response prevents any strong inhomogeneity on scales larger than
the Planck scale from occurring.
The magnitude of fluctations on the Planck scale can be estimated as follows.

If the handle multiplicity is large, then the standard deviation of
statistical fluctuations $\sigma$ decrease with $N_{\rm BHll}$ as
$\sigma (m) = 1/(2N_{\rm BHll}(m))^{1/2}$. Given the discrete
nature of induced mini BHs, shot noise is the proper statistic for this
phenomenon.

Also, the typical mass of a longer-lived mini BH was estimated above as
$m_{\rm BHll}\sim 10^3m_P$, which yields $l\sim 10^3 l_P$. Once the epoch
of primordial longer-lived BH formation commences,
$\Lambda (m)=2N_{\rm BHll}m_P/l^3$ is driven to the Planck energy density.
This constitutes a form of topological reheating, because it is determined by
the number of (quantum mechanically leaky) event horizons in the universe.
It also implies that $N_{\rm BHll}\sim 10^9$ for $l\sim 10^3 l_P$.

The latter number of $10^9$ should be close to the maximum value and thus
set $\sigma$. An energy density on the order of the Planck density strongly
suppresses the evaporation of induced mini BHs, thereby increasing the number
of longer-lived BHs $N_{\rm BHll}$. However, the accretion rate of already
existing longer-lived BHs would also be very high at the the Planck density.
The latter boosts the mass of these BHs, and thus their longevity since
$t_e\propto m_{\rm BH}^3$.

Hence, $N_{\rm BHll}\sim 10^9$ is not likely to
grow much before a BH exists with a life span that exceeds the age of the
universe at that time, thus freezing in the value of $L_f$.
In all, one finds a very acceptable value of $\sigma\lesssim 2\times 10^{-5}$
for the magnitude of cosmological mass-energy fluctuations.
Given that $N_{\rm BHll}(m)>>1$, the Poisson distribution of shot noise will
typically be indistinguishable from Gaussian noise.

These perturbations are a result of the discrete nature of $\Lambda (m)$
and find their place in $T_{\mu\nu}$, so on the right hand side of the
Einstein equation. Indeed, the very fact that GR relates geometry and
matter allows one to include the global topological impact of dark energy
as well as its local realization in energy-momentum.

\subsection{Black Hole X: The End of Inflation}

When $L_f$ is frozen in, this is locally determined by the favorable accretion
history of one particular BH. However, $L_f$ is a global quantity. So if it is
much larger than $l$, then the dark energy density drops very strongly and
the universe exits inflation in a homogeneous (so-called graceful) manner.
The actual size $l_X$ of the longer-lived mini BH $X$ that sets $L_f$ requires
a detailed (numerical) computation.
Nevertheless, a rough estimate of $l_X$ can be obtained.

Typically, $l\sim 10^3 l_P$ while $t_e\propto m_{\rm BH}^3$ strongly depends
on BH mass. Therefore, a relatively modest increase in the longer-lived mini
BH mass likely suffices to freeze in $L_f$, given that the age of the universe
at the onset of inflation is less than $\sim 10^9t_P$ and set by $t_e$.
So, taking an excursion from the mean by a factor of 10, yields
$m_X\sim 0.03$ g and an evaporation time $t_X\sim 10^{12} t_P$ for the first
BH that lives longer than the age of the universe at that time.

Obviously, $l_X\sim 10^{4} l_P$ is very much smaller than the size $L_f$ of
the universe at that time.
In fact, $L_f/l_X\sim 10^{43}$ while $N_{\rm BHll}\sim 10^9$ during inflation
and $N_{\rm BH}\sim 10^{19}$ today. Assuming that $l_X$ is an upper limit,
this yields a ratio between the current dark energy density and the
inflationary vacuum energy density of
$\zeta\sim (N_{\rm BH}/N_{\rm BHll})(l_X/L_f)^3\lesssim 10^{-119}$.
This is approximately equal to the well known 120 orders of magnitude vacuum
energy gap, and alleviates it through the MP and the EMP.

In all, the vacuum energy density at the end of the inflation diminishes
{\it globally} by many orders of magnitude when BH X forms. This is a very
desirable feature to match the very early universe energy scale with the
current one.
At the beginning of inflation the ambient radiation temperature was high
and mini BH evaporation surpressed. The MP induces mini BHs every Planck
time, providing fresh candidates. Combined this at least helps to build up a
tail of (lucky) massive longer-lived BHs that can survive over cosmic
times.\footnote{These BHs may constitute a form of dark matter when more
massive than $\sim 10^{15}$ g. If driven by self-multiplication of order
$1+\mu_1$ or $(1+\mu_1)^2$ then a natural square root mean dark-to-baryonic
matter ratio is $\sim [(1+\mu_1)^3]^{1/2}=(27)^{1/2}\approx 5.2$.
Conversely, such dark matter can decay due to Hawking evaporation, comfortably
producing particles of more than several GeV, for $\lesssim 10^{14}$ g.}
That said, BH $X$ may not have lived to the present day. Still, that is
irrelevant. BH $X$ is forever a part of the universe's 4-space past.

\subsection{The Scaling of Cosmological Perturbations}

The longer-lived mini BHs are the driving force behind inflation in this work.
Consequently, it is to be expected that the first and second order time
derivative of $N_{\rm BHll}$ is small. Afterall, accretion does not affect BH
number, while the evaporation time of these BHs is required to be at least on
the order of the age of the universe in the first place.
Because $N_{\rm BHll}$ acts as the single field here, there is thus a
strong lack of rapidity in this type of inflation.
The above fits the picture of slow-roll inflation.

The very nature of mini BH driven inflation avoids any problems with
entropy generation
(or reheating). In fact, the evaporation of induced mini BHs, in
every Planckian volume, optimally uses (the second law of) thermodynamics
at very large $T_{\rm BH}$.
This while Eq.~(22) supports an exponential dependence for $n_1(m)$ with
topological time $m$.
It is also good to recall the comments on $L_f$ and $\tau_f$ here, and to
realize that our universe is properly embedded geometrically in 4-space only
when BH $X$ forms and inflation ends. The latter can be pursued a bit further
to also constrain the scaling of cosmological perturbations.

The induced mini BHs, denoted by $\Delta N$, appear in a quantum space-time
$L(T^3)^+$ that enjoys an exponentially rapid increase in geometric size
$R(t)$.
This picture introduces temporal effects, even when $N_{\rm BHll}$ is
intrinsically constant, and can be used to determine the tensor-to-scalar
ratio $r$ and the spectral index $n_s$, as follows.

Consider a local Planckian observer during inflation who is comoving with
$R(t)$, so conformally, while $R(\tau )=l_P\tau$ measures the size of
the universe in topological time $m>>1$ and $\Delta m>>1$, since the onset
of inflation. Then $t\propto ln(R)\propto ln(\tau )$ and $dt=d\tau /\tau$.
If $2\Delta m<m$ does not hold, then the coarse-grained nature of topological
time affects the appearance of the induced mini BHs $\Delta N$ on $L(T^3)^+$
because time slices $m$ are topologically identified over $\Delta m>>1$.

Under the assumption that the longer-lived BHs are local perturbations only,
one has that $L(T^3)^+\approx L(T^3)$.
The local Planckian observer, measuring for a geometric time interval $dt$
(not necessarily a small number), then notes the following.
Induced mini BHs are spread out through time over intervals $dt>>t_P$.
This generates very specific curvature perturbations. The observer sees
mini BHs appear and disappear with a duty cycle that is set by the
topological time $\tau$, for $\Delta m\rightarrow m$ and
$\delta_t = 1/m$ in $T^3|[R,\delta_t^{-1}]\in L(T^3)$.

Furthermore, the scale invariance of these perturbations is broken because the
duration $\tau$ of inflation sets a characteristic size $R(\tau )/l_P$ on
$L(T^3)$. So, even though $N_{\rm BHll}$ is constant, one has the situation
that there is a gravitational wave generating quantity $r$ that should obey

\begin{equation}
r=\frac{dt}{t_P}\ \frac{d(\Delta N)}{d\tau}.
\end{equation}

The observer concludes that there are ${\rm n}=\tau$
e-foldings ($\tau\rightarrow\tau +1$ stretches $t$ by a factor $e$)
that diminish this tensor component during inflation. Indeed,

\begin{equation}
\frac{d(\Delta N)}{d\tau}= t_P\frac{d(\Delta N)}{\tau dt}\propto\frac{1}{{\rm n}}.
\end{equation}

Also, every induced mini BH is attached to a heptaplet of three-tori on
$L(T^3)$ every $t_P$, driving its 7-fold topological identification across
$\Delta m$. The observer counts these as 7 in his local
geometric time frame, but $t\propto ln(\tau )$ and this yields
$d(\Delta N)=ln(7)$. Hence, combined this gives $r=ln(7)/{\rm n}\approx 0.036$.
According to the Planck results in Ref.~42, $r<0.11$ and one finds comfortable
consistency.

In the constant $N_{\rm BHll}$ approximation, $r$ is fully driven by
the distinction between topological and geometric time intervals.
This while the spatial scaling properties of $T^3|[R,\delta_t^{-1}]\in L(T^3)$
are completely expressed by the difference between $R(\tau )$ and $R(t)$, with
the latter pertaining to our observational perspective.
From above it also follows that $r\propto 1/\tau \propto 1/R(\tau )$,
which is scale free.

So to find the relation between $r$ and $n_s$ one realizes that there is no
fundamental topological distinction between $\tau$ and $R(\tau )$ or $t$ and
$R(t)$.
Any member of these two pairs can be taken as a temporal or spatial variable.
Therefore, $r$ is equal to the relative amplitude of curvature
perturbations ($\tau$ versus $t$)
associated with mini BH induction on a dynamic $L(T^3)$ background
{\it as well as} to the deviation ($R(\tau )$ versus $R(t)$) from scale-free
behavior of that $L(T^3)$ background.
Subsequently, $r$ and $n_s$ are equivalent and related by
$n_s=1-r\approx 0.964$. This agrees very well with recent Planck
observations$^{42}$, which indicate $n_s=0.9603\pm 0.0073$.

In standard inflation theory, there is the subtle point of when modes cross
the horizon. Such a phenomenon is not pertinent in inflation driven by
topological dynamics.
Rather, the Gaussian statistics of $T_{\mu\nu}$ in terms of $\sigma (m)$
are spatially uniform in a global manner thanks to the MP and EMP.
With the additional numbers $n_s$ and $r$,
the energy-momentum tensor has its properties fully specified.

\section{Closing Remarks}

\subsection{Detecting the Topological to Geometric Time Transition}

An additional route to explore the global character of our universe
is to look for a macroscopic manifestation of the Planckian lattice
of three-tori.
The structure $L(T^3)^+(m)$ is being build up further as we speak,
rendering it exponentially more fine-grained spatially through Eq.~(21).
This precludes the occurrence of a multiply connected space-time on large
spatial scales for cosmologically short time intervals.\footnote{For
$R(t)<ct$, like in the absence of an inflationary period, spatial
identifications of $T^3|[R,\delta_t^{-1}]$ can become
visible to local observers.}
However, the transition of topological time $m$ to
geometric time $t$ was described in section 4.2. The validity of the
latter only holds for time intervals $\Delta m$ that obey $2\Delta m<m$.

An interesting situation then occurs for the cosmic microwave background
(CMB). The current age of the universe is $t_0\approx 10^{61}t_P=m_0t_P$,
about 14 billion years.
The CMB has been travelling towards us since very close to that time,
because the last scattering surface occurs only $\sim 10^{5.5}$ years after
the big bang. Hence, geometric time is not a proper variable to describe
topological dynamics cosmologically.

In particular, $L_0=ct_0$ is the largest scale that can be observed today,
whatever the size $R(t_0)>L_0$ of the universe. On $L_0$, the only
observable constituent of $L(T^3)^+$ is a pair of three-tori.
For $2\Delta m\ge m_0$, the members of this pair are macroscopically separated
and topologically identified on a scale $L_0/2$. Through this $T^3$ connection
CMB radiation is temporally displaced over $t_0/2$.
Because $L_0$ is what an observer perceives of the universe at $t_0$,
geometric time only holds for scales $2L<L_0$. Note here that quantum
space-time, so $L(T^3)^+(m)$, is still continuous. It is our perception
thereof that becomes crude.

This coarse-grained character of geometric time therefore leads to global
spatial asymmetries in the CMB's tiny temperature fluctuations.
This topological redistribution process has some efficacy, which can be
estimated as follows.
As a consequence of the MP, each three-torus has a multiplication factor
$F=1+\mu_0=7$, i.e., two heptaplets of three-tori perturb the CMB. Hence,
some amount of radiation $U$ on the scale $L_0/2$ of one three-torus is
temporally displaced across a scale $L_0/2$ to the other three-torus, and
vice versa. One thus finds for the amplitude $B=dU/U$ of the resulting
scalar fluctuation that

\begin{equation}
B=\frac{1}{2F}\approx 0.07.
\end{equation}

Recent Planck data$^{44}$ indicate a large scale asymmetry with an amplitude
of $B=0.07\pm 0.02$. This empirical value of $B$ is suggestively close to the
theoretical prediction. Furthermore, the Planck data indicate multipole
scales $\lambda\lesssim 60$. These correpond to spatial scales of $L_0-L_0/60$,
fully consistent with $L_0/2$. This while such a $L(T^3)^+$ driven asymmetry
does not complicate the inflationary scenario discussed above.

Electromagnetic signals across time intervals $\Delta m <m_0/2$ experience
no temporal displacement effects. But the early universe, roughly $z>0.6$ for
more than 7 billion years ago, should also exhibit temporal displacement
features. The distribution of luminous matter (e.g., galaxies) is the best
place to look for these fluctuations. In a series of time slices through
4-space, they constitute a violation of the cosmological principle, driven
by coarse-grained time as an expression of quantum space-time topology.
Of course, the small amplitude $B$ of about 7\% makes such temporal
displacement fluctuations on a scale $L_0/2$ difficult to discern in an
ensemble of large scale tracers.

\subsection{Identification by Multiplication}

Suppose the predictions on dark energy and inflation in this work are
confirmed. Then quantum gravity, as a topological extension of GR in the
form of Eq.~(17), is an expression of the global multiply connected nature of
Planckian space-time. In this, the EMP provides the link between the
macroscopic and microscopic cosmos. Long-lived BHs induce mini BHs on
the Planck scale, as miniature copies of the large scale originals.

The work presented here advocates the role of a very specific space-time
topology for quantum gravity. In it, homeomorphically distinct paths in
the form of loops take central stage and the prime manifolds $T^3$ and
$S^1\times S^2$ express Feynman's superposition and Wheeler's quantum foam,
respectively. The fundamental principle is the MP. The resulting quantum
space-time is $L(T^3)^+(m)$.

The MP dictates that the paths of quantum space-time are dynamical.
Nature, so all observers combined, counts distinct paths unremittingly in
order to identify them. Because it takes one to know one, a path that has
been identified on the Planck scale must, in response to such observation,
also have been multiplied as a part of quantum space-time.
In all, the distinct paths that drive topological fluctuations on the Planck
scale and underlie quantum gravity, obey a simple rule: multiply to identify.

Recall that Eq.~(17) only appears simple. All the
properties of Einstein gravity, as well as its possible
modifications/unifications, are both part of and reduced to the complex
temporal evolution of the number of quantum mechanically leaky event horizons
on the Planck scale.

Overall, Eq.~(17) has pertinent consequences
for dark energy, inflation and the nature of time.
A more complete exploration of the solutions to Eq.~(17) may bring out new
features for our cosmic evolution. In fact, such solutions can also help to
test different modifications of GR against the observed history of
our universe.

Topologically extended GR, expressed in discrete global time $m$ (through,
e.g., a foliation based on BH longevity) and continuous intervals
$(m+)=(m,m+1)$, is
\begin{equation}
G_{\mu\nu}[{\rm n}]+\Lambda_A (m) g_{\mu\nu}=8\pi GT_{\mu\nu}[n_1(m),\sigma (m)]_{r,n_s},
\end{equation}
with $A=L_f$ or $A=R_{\rm Sll}$ (both generally covariant),
$\Lambda (m)/m_P=2n_1(m)/A^3$ and $\sigma (m) = 1/(2N_{\rm BHll}(m))^{1/2}$
the standard deviation of Gaussian mass-energy fluctuations induced by
longer-lived BHs in the early universe, for ${\rm n}\approx 55$ e-foldings,
$r=ln(7)/{\rm n}\approx 0.036$ and $n_s=1-r\approx 0.964$. In this,
\begin{equation}
n_1(m)[\{M_{k'}\}] = n_1(m)[\{M_k\}] + 2n_1(m)[m_P] + \Delta F(m+)-\Delta E(m+)-\Delta M(m+),
\end{equation}
for the masses $M_k$ ($M_{k'}$) that BHs $k=1...n_1(m)$ ($k'=1...n_1(m+1)$)
have, and for pairs of induced BHs with mass $m_P$.
Periodic boundary conditions are imposed by the initial ($m=0$) three-torus
that leads to $L(T^3)$ and quantum superposition.
To solve quantum gravity in the form of Eqs.~(41) and (42) it is crucial
that the evolution of the BH ensemble $n_1$ in $\Lambda$ and $T_{\mu\nu}$
is followed completely.

\section*{Acknowledgments}

The author is very greatful to Elly van 't Hof for stimulating discussions
on the nature of quantum space-time and its relation to every-day life.
The author is also greatful to Pieter de Laat for stimulating discussions
on the roots of quantum physics.
The author thanks Aycin Aykutalp for help with figure 1 and interesting
discussions.
Finally, the author thanks the organizers and participants during the 7th
spontaneous workshop on hot topics in modern cosmology (Carg\`ese,
6-11 May 2013), specifically Roland Triay,
Brett Bochner, Carlos Romero, Lorenzo Reverberi, Mario Novello,
Grigory Rubtsov, Denis Comelli, and Dmitry Gorbunov.



\begin{thebibliography}{0}    


\bibitem{} A. Einstein, {\it Preuss. Akad. Wiss. Berlin} {\bf Sitzber.} (1915) 778.

\bibitem{} A. Einstein, {\it Letter to Arnold Ehrenfest on his discovery of GR} (28 Nov. 1915).

\bibitem{} J.A. Wheeler, {\it Ann. of Phys.} {\bf 2} (1957) 604.

\bibitem{} J.A. Wheeler, {\it Quantum Cosmology}, eds. L.Z. Fang \& R. Ruffini  (World Scientific, 1987) p.27-92. This book provides a very nice overview of earlier developments in the field.

\bibitem{} B.S. DeWitt, {\it The Physical Review} {\bf 160} (1967) 1587.

\bibitem{} A. Ashtekar, {\it Phys. Rev. D} {\bf 36} (1987) 1587.

\bibitem{} A. Ashtekar, {\it Loop Quantum Gravity and the Planck Regime of Cosmology}, arXiv:1303.4989 [gr-qc]. This paper contains many references to interesting works on loop quantum gravity cosmology.

\bibitem{} S. Kaku, {\it Introduction to Superstrings}, 2nd edn. (Springer-Verlag, 1990).

\bibitem{} R. Dijkgraaf and C. Vafa, {\it Nuc. Phys. B} {\bf 644} (2002) 3.

\bibitem{} J.M. Maldacena, {\it AdTMP} {\bf 2} (1998) 231.

\bibitem{} L. Smolin, {\it General relativity as the equation of state of spin foam}, arXiv:1205.5529 [gr-qc].

\bibitem{} M. Spaans, {\it Nuc. Phys. B} {\bf 492} (1997) 526.

\bibitem{} M. Spaans, {\it J. Phys.: Conf. Ser.} {\bf 410} (2013) 012149.

\bibitem{} L. Smolin, {\it The case for background independence}, arXiv:hep-th/0507235.

\bibitem{} S.W. Hawking, {\it Comm. Math. Phys.} {\bf 43} (1975) 199 as well as S.W. Hawking and D.N. Page, {\it ApJ} {\bf 206} (1976) 1.

\bibitem{} K.S. Virbhadra and G.F.R. Ellis, {\it Phys. Rev. D} {\bf 65} (2002) 103004 as well as K.S. Virbhadra and C.R. Keeton, {\it Phys. Rev. D} {\bf 77} (2008) 124014

\bibitem{} A.G. Riess {\it et al.}, {\it Astron. J} {\bf 116} (1998) 116.

\bibitem{} S. Perlmutter {\it et al.}, {\it ApJ} {\bf 517} (1999) 565.

\bibitem{} B.P. Schmidt {\it et al.}, {\it ApJ} {\bf 507} (1998) 46.

\bibitem{} R. Amanullah {\it et al.}, {\it ApJ} {\bf 716} (2010) 712.

\bibitem{} P.M. Garnavich {\it et al.} {\it ApJ} {\bf 509} (1998) 74.

\bibitem{} R.P. Feynman, {\it Rev. Mod. Phys.} {\bf 20} (1948) 367.

\bibitem{} M. Gromow and B. Larson, {\it Publ. Math.} {\it 58} (1976) 295.

\bibitem{} J. Hempel, {\it 3-Manifolds; Annals of Mathematical Studies}, Vol. 86 (Princeton, 1976).

\bibitem{} M. Nakahara, {\it Geometry, Topology and Physics}, 1st edn. (Adam Hilger, 1990).

\bibitem{} S. D\"urr, T. Nonn and G. Rempe, {\it Nature} {\bf 395} (1998) 33.

\bibitem{} A.M. Hopkins and J.F. Beacom 2006, {\it ApJ} {\bf 651} (2006) 142.

\bibitem{} A. Aykutalp and M. Spaans, {\it Testing the Proposed Connection between Dark Energy and Black Holes}, arXiv:1201.6372 [astro-ph.CO].

\bibitem{} G. Hinshaw {\it et al.}, {\it Nine-Year Wilkinson Microwave Anisotropy Probe (WMAP) Observations: Cosmological Parameter Results}, arXiv:1212.5226 [astro-ph.CO].

\bibitem{} P.A.R. Ade {\it et al.}, {\it Planck 2013 results. XVI. Cosmological parameters}, arXiv:1303.5076 [astro-ph.CO].

\bibitem{} R. Anantua, R. Easther and J.T. Giblin, {\it Phys. Rev. Lett.} {\bf 103} (2009) 111303.

\bibitem{} E. Komatsu {\it et al.}, {\it ApJS} {\bf 192} (2011) 18.

\bibitem{} M. Sullivan {\it et al.} {\it ApJ} {\bf 737} (2011) 102.

\bibitem{} K. Bamba, S. Capozziello, S. Nojiri and S.D. Odintsov, {\it Astrophysics and Space Science} {\bf 342} (2012) 155. Also, there is S. Tsujikawa, {\it Quintessence: A Review}, arXiv:1304.1961 [gr-qc]. These papers contain many references to different dark energy approaches.

\bibitem{} S. Nojiri and S.D. Odintsov, {\it Physics Reports} {\bf 505} (2011) 59. Also, there is T.P. Sotiriou and V. Faraoni, {\it Rev. Mod. Phys.} {\bf 82} (2010) 451.

\bibitem{} A.H. Guth, {\it Phys. Rev. D} {\bf 23} (1981) 347.

\bibitem{} A.D. Linde, {\it Phys. Lett. B} {\bf 108} (1982) 389.

\bibitem{} A. Albrecht and P.J. Steinhardt, {\it Phys. Rev. Lett} {\bf 48} (1982) 1220.

\bibitem{} A.A. Starobinsky, {\it Phys. Lett. B} {\bf 91} (1980) 99.

\bibitem{} A.A. Starobinsky, {\it JETP Lett.} {\bf 55} (1992) 489.

\bibitem{} S. Hocuk and M. Spaans, {\it A\&A} {\bf 536} (2011) A41.

\bibitem{} P.A.R. Ade {\it et al.}, {\it Planck 2013 results. XII. Constraints on inflation}, arXiv:1303.5082 [astro-ph.CO].

\bibitem{} A. Ijjas, P.J. Steinhardt and A. Loeb, {\it Inflationary paradigm in trouble after Planck2013}, arXiv:1304.2785 [astro-ph.CO].

\bibitem{} P.A.R. Ade {\it et al.}, {\it Planck 2013 results. XIII Isotropy and statistics of the CMB}, arXiv:1303.5083 [astro-ph.CO].

\end{thebibliography}
\end{document}